\documentclass[twocolumn,pra,showpacs]{revtex4}

\usepackage{amsmath}
\usepackage{graphicx}
\usepackage{amsfonts}
\usepackage{amssymb}

\begin{document}

\title{Quantum state transfer between fields and atoms\\
in Electromagnetically Induced Transparency}
\author{A. Dantan\footnote[3]{email: dantan@spectro.jussieu.fr}}

\author{M. Pinard}

\affiliation{Laboratoire Kastler Brossel, Universit\'{e} Pierre et
Marie Curie\\
Case 74, 4 place Jussieu, 75252 Paris Cedex 05, France}

\date{\today}

\newcommand{\beq}{\begin{equation}}
\newcommand{\eeq}{\end{equation}}
\newcommand{\beqr}{\begin{eqnarray}}
\newcommand{\eeqr}{\end{eqnarray}}
\newcommand{\lb}[1]{\label{#1}}
\newcommand{\ct}[1]{\cite{#1}}
\newcommand{\dpr}{\delta P_r}
\newcommand{\dpd}{\delta P_2}
\newcommand{\da}{\delta A_2}
\newcommand{\dain}{\delta A_2^{in}}
\newcommand{\djx}{\delta J_x}
\newcommand{\djy}{\delta J_y}
\newcommand{\dsx}{\delta S_x^{in}}
\newcommand{\dsy}{\delta S_y^{in}}

\begin{abstract}
We show that a quasi-perfect quantum state transfer between an
atomic ensemble and fields in an optical cavity can be achieved in
Electromagnetically Induced Transparency (EIT). A squeezed
vacuum field state can be mapped onto the long-lived atomic spin
associated to the ground state sublevels of the $\Lambda$-type
atoms considered. The EIT on-resonance situation show interesting similarities
with the Raman off-resonant configuration. We then show how to
transfer the atomic squeezing back to the field exiting the cavity,
thus realizing a quantum memory-type operation.
\end{abstract}

\maketitle

\section{\bigskip Introduction}

If photons are known to be fast and robust carriers of quantum
information, a major difficulty is to store their quantum state.
In order to realize scalable quantum networks \ct{divincenzo}
quantum memory elements are required to store and retrieve photon
states. To this end atomic ensembles have been widely studied as
potential quantum memories \ct{lukin,polzik}. Indeed, the
long-lived collective spin of an atomic ensemble with two ground state
sublevels appears as a good candidate for
the storage and manipulation of quantum information conveyed by
light \ct{duan}. Various schemes have been studied:
first, the recent "slow-" and "stopped-light"
experiments have shown that it was possible to store a light pulse
inside an atomic cloud \ct{hau,phillips} in the
Electromagnetically Induced Transparency (EIT) configuration
\ct{harris}. EIT is known to occur when two fields are both one-
and two-photon resonant with 3-level $\Lambda$-type atoms, which
allows one field to propagate without dissipation through the
medium. However, the storage has only been demonstrated for
classical variables so far.\\
On the other hand, the stationary
mapping of a quantum state of light (squeezed vacuum)
onto an atomic ensemble has been experimentally demonstrated, this time in an
off-resonant Raman configuration \ct{polzik2} and in a single pass scheme. Squeezing transfer from
light to atoms is also interesting in relation to "spin squeezing"
\ct{wineland} and has been widely studied \ct{polzik3,kuzmich,kozhekin,vernac1,molmer,dantan1}.\\
In this paper, unlike the single-pass approaches, we consider a cavity configuration, allowing a full quantum
treatment of the fluctuations for the atom-field system \ct{vernac1}. We show
that it is possible to continuously transfer squeezing, either in an EIT or Raman
configuration, between a cloud of cold 3-level $\Lambda$-type
atoms placed in an optical cavity and interacting with two fields:
a coherent pump field and a broadband squeezed vacuum field.\\
The paper is organized as follows: Sec. \ref{model} briefly describes the system,
in Sec. \ref{lfapprox} we develop a simplified model and study the conditions under
which the squeezing transfer is optimal. Both EIT and Raman schemes result in a
quasi-perfect transfer, which is not true for an arbitrary detuning.
In Sec. \ref{full} we check that these conclusions are in agreement with full
quantum calculations, evaluate the transfer robustness with respect to a detuning from two-photon
resonance and generalize to the case of non-zero amplitude fields. Last, we present
a simple readout scheme for the atomic squeezing in Sec. \ref{lecture}: the squeezing stored in the atomic medium
can be retrieved on the vacuum field exiting the cavity by switching off and on the pump field. The efficiency of the readout
process is conditioned by the temporal profile of the local oscillator used to detect the outgoing vacuum field fluctuations,
and can be close to 100\% by an adequate choice of the local oscillator profile.

\section{Model system}\lb{model}

\subsection{Atom-fields evolution equations}

The system considered in this paper is a set of $N$ $3$-level
atoms in a $\Lambda $ configuration, as represented in Fig 1. On
each transition $i\longrightarrow 3$ the atoms interact with one
mode of the electromagnetic field, $A_{i}$ in an optical cavity
($i=1,2$). The detunings from atomic resonance are $\Delta _{i}$
and the cavity detunings $\Delta _{ci}$. The $3$-level system is
described using $9$ collective operators for the $N$ atoms of the
ensemble: the populations $\Pi _{i}=\sum\limits_{\mu =1}^{N}\left|
i\right\rangle _{\mu }\left\langle i\right| _{\mu }$ ($i=1-3)$,
the components of the optical dipoles $P_{i}$ in the frames
rotating at the frequency of their corresponding lasers and their
hermitian conjugates and the components of the dipole associated
to the ground state coherence: $ P_{r}=\sum\limits_{\mu
=1}^{N}\left| 2\right\rangle _{\mu }\left\langle 1\right| _{\mu }$
and $P_{r}^{\dagger }$.
\begin{figure}[h]
  \centering
  \includegraphics[width=5cm]{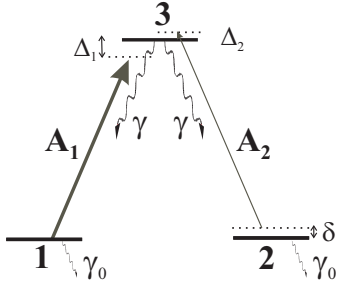}
  \caption{Three-level system in a $\Lambda$ configuration.}\label{fig1}
\end{figure}
The atom-field coupling constants are defined by $g_i={\cal
E}_{0i}d_i/\hbar $, where $d_i$ are the atomic dipoles, and ${\cal
E}_{0i}=\sqrt{ \hbar \omega _{i}/2\epsilon _{0}{\cal S}c}$. With
this definition, the mean square value of a field is expressed in
number of photons per second. To simplify, the decay constants of
dipoles $P_{1}$ and $P_{2}$ are both equal to $\gamma$. In order
to take into account the finite lifetime of the two ground state
sublevels $1$ and $2$, we include in the model another decay rate
$\gamma _{0}$, which is supposed to be much smaller than $\gamma
$. Typically, the atoms fall out of the interaction area with the
light beam in a time of the order of a few millisecond, whereas
$\gamma $ is of the order of a few MHz for excited states. We also
consider that the sublevels $1$ and $2$ are repopulated with
in-terms $\Lambda _{1}$ and $\Lambda _{2}$, so that the total
atomic population is kept constantly equal to $N$.

The system evolution is given by a set of quantum
Heisenberg-Langevin equations

\begin{eqnarray}
\dot{\Pi}_{1} &=&ig_1A _{1}^{\dagger}P_{1}-ig_1A
_{1}P_{1}^{\dagger }+\gamma \Pi _{3}-\gamma
_{0}\Pi _{1}+\Lambda _{1}+F_{11}  \label{pi1} \nonumber\\
\dot{\Pi}_{2} &=&ig_2A_{2}^{\dagger }P_{2}-igA_{2}P_{2}^{\dagger
}+\gamma \Pi _{3}-\gamma _{0}\Pi _{2}+\Lambda
_{2}+F_{22}  \label{pi2} \nonumber\\
\dot{\Pi}_{3} &=&-(ig_1A _{1}^{\dagger
}P_{1}-ig_1A_1P_{1}^{\dagger })-(ig_2A_{2}^{\dagger
}P_{2}-ig_2A_{2}P_{2}^{\dagger
})\nonumber\\
&&\hspace{3cm}-2\gamma\Pi _{3}+F_{33}  \label{pi3} \nonumber\\
\dot{P}_{1} &=&-(\gamma +i\Delta _{1})P_{1}+ig_1A_1(\Pi
_{1}-\Pi _{3})+ig_2A_{2}P_{r}^{\dagger }+F_{1}  \label{p13} \nonumber\\
\dot{P}_{2} &=&-(\gamma +i\Delta _{2})P_{2}+ig_2A_{2}(\Pi _{2}-\Pi
_{3})+ig_1A_1 P_{r}+F_{2}  \label{p23} \nonumber\\
\dot{P}_r&=&-(\gamma_0-i\delta)P_r+ig_1A_1^{\dagger}P_2-ig_2A_2P_1^{\dagger}+f_{r}\label{pr}\nonumber\\
\dot{A}_{1} &=&-(\kappa +i\Delta _{c1})A_{1}+\frac{ig_1}{\tau }
P_{1}+\sqrt{\frac{2\kappa }{\tau
}}A_{1}^{in}  \label{A1}\nonumber\\
\dot{A}_{2} &=&-(\kappa +i\Delta _{c2})A_{2}+\frac{ig_2}{\tau }
P_{2}+\sqrt{\frac{2\kappa }{\tau }}A_{2}^{in}  \nonumber\label{A2}
\end{eqnarray}

where $g_{1,2}$ are assumed real, $\delta =\Delta _{1}-\Delta
_{2}$ is the two-photon detuning, $\kappa $ is the intracavity
field decay and $\tau $ the round trip time in the cavity. The
$F$'s are standard $\delta$-correlated Langevin operators
taking into account the coupling with the other cavity modes. From
the previous set of equations, it is possible to derive the steady
state values and the correlation matrix for the fluctuations of
the atom-fields system (see e.g. \cite{vernac1}).

\subsection{Decoupled equations for the fluctuations}

In the case $\langle A_2^{in}\rangle=0$ and $\Lambda_2=N\gamma_0$,
all the atoms are pumped in $|2\rangle$, so that only
$\langle\pi_2\rangle$ is non zero in steady state. Here, we assume
that $\langle A_2\rangle$ is zero, even if the number of
intracavity photons is non-zero \textit{stricto sensu} for a
squeezed vacuum, this assumption is valid as long as the number of
intracavity photons is much smaller than the number of atoms. In
this case, the fluctuations for $\delta P_r$, $\delta P_2$ and
$\delta A_2$ are then decoupled from the other operators
fluctuations
\beqr \dot{\dpr}&=&-(\gamma_0-i\delta)\dpr+i\Omega \dpd+f_r\lb{dpr}\\
\dot{\dpd}&=&-(\gamma+i\Delta)\dpd+i\Omega \dpr+igN\da+F_{2}\lb{dpd}\\
\dot{\da}&=&-(\kappa+i\Delta_c)\da+\frac{ig}{\tau}\dpd+\sqrt{\frac{2\kappa}{\tau}}\dain\lb{da}\eeqr
To simplify, we omit the subscript $2$ for $g$, $\Delta$ and
$\Delta_c$, and assume that the Rabi pulsation associated to the
pump field $\Omega=g_1\langle A_1\rangle$ is real. The atomic spin
associated to the ground states is aligned along $z$ at steady state: $\langle
J_z\rangle=\langle \pi_2-\pi_1\rangle/2=N/2$. We will place
ourselves in this situation, which not only allows for analytical
calculations and provides simple physical interpretations, but can
also be generalized to arbitrary states for fields $A_1$ and
$A_2$, as we will show further.\\
To characterize the quantum state of the atomic ensemble we look
at the fluctuations of the spin components in the plane orthogonal
to the mean spin: $J_x=(P_r+P_r^{\dagger})/2$ and
$J_y=(P_r-P_r^{\dagger})/2i$. The spin component $J_{\theta}=J_x\cos\theta
+J_y\sin\theta$ in the ($x,y$)-plane is said to be spin-squeezed when its
variance is less than the coherent state value $|\langle J_z\rangle|/2$, and
the degree of spin-squeezing is
given by \ct{ueda}
\beqr \Delta J^2_{min}=\min_{\theta}\frac{\Delta
J^2_{\theta}}{|\langle J_z\rangle|/2}<1\eeqr

\section{Adiabatical eliminations in the low frequency limit}\lb{lfapprox}

\subsection{EIT configuration}

Since the ground state sublevels have a long lifetime compared to
the excited state ($\gamma_0\ll\gamma$), and in the bad cavity
limit ($\kappa\gtrsim\gamma$), the atomic spin associated to
levels 1 and 2 evolves much slowly than the field or the optical
coherence. Fourier-transforming Eqs.
(\ref{dpr}-\ref{dpd}-\ref{da}) and adiabatically eliminating
$\dpd$ and $\da$, one gets a simplified equation for the
ground state coherence fluctuations
\beqr
\left[\gamma_0-i\delta+\frac{\Omega^2(\kappa+i\Delta_c)}{d}-i\omega\right]\dpr(\omega)=&\lb{generalLF}\\
\nonumber-\frac{gN\Omega}{d}\sqrt{\frac{2\kappa}{\tau}}\dain(\omega)+\frac{i\Omega(\kappa+i\Delta_c)}{d}F_{2}(\omega)+& f_r(\omega)\\
\nonumber\text{with}\hspace{1cm}d=(\kappa+i\Delta_c)(\gamma+i\Delta)+\frac{g^2N}{\tau}&\eeqr
In the so-called EIT configuration, the fields are one- and
two-photon resonant: $\Delta=\delta=0$. Moreover, for the
squeezing transfer to be optimal, one must have a zero-cavity
detuning: $\Delta_c=0$ \ct{vernac1,dantan1}. The equations for the
spin components in the ($x,y$)-plane are then
\beqr
(\tilde{\gamma}_0-i\omega)\djx &=&\frac{-gN\Omega}{\gamma(1+2C)\sqrt{T}}\delta A_p^{in}+\tilde{f}_x\lb{jxeit}\\
(\tilde{\gamma}_0-i\omega)\djy
&=&\frac{-gN\Omega}{\gamma(1+2C)\sqrt{T}}\delta A_q^{in}+\tilde{f}_y\lb{jyeit}
\eeqr
with an effective decay constant
$\tilde{\gamma_0}=\gamma_0+\frac{\Gamma_E}{1+2C}$,
$\Gamma_E=\Omega^2/\gamma$ being the one-photon resonant pumping rate.
$T=2\kappa\tau$ is the coupling mirror transmission of the single-input cavity and
$\tilde{f}_x$, $\tilde{f}_y$ are effective Langevin
operators
\beqr \tilde{f}_{x}= f_{x}-\frac{\Omega}{\gamma(1+2C)}F_{y},
\hspace{0.3cm}\tilde{f}_{y}=
f_{y}+\frac{\Omega}{\gamma(1+2C)}F_{x}\lb{feit}\eeqr
$A_p=A_2+A_2^{\dagger}$ and $A_q=i(A_2^{\dagger}-A_2)$ are the standard amplitude and phase
quadratures for the squeezed vacuum field. Although the two modes $A_{1,2}$ do not need to be orthogonally
polarized modes, it is rather convenient for the discussion to
consider them as $\sigma_+$ and $\sigma_-$ modes of the field.
In order to stress the similarity between the atomic spin and the
Stokes vector which characterize the polarization state of the
light, we introduce
\beqr
S_0&=&\hspace{0.4cm}A_1^{\dagger}A_1+A_2^{\dagger}A_2,\hspace{0.3cm}S_y=i(A_1^{\dagger}A_2-A_1A_2^{\dagger})\nonumber\\
S_x&=&-(A_1^{\dagger}A_2+A_1A_2^{\dagger}),\hspace{0.2cm}S_z=A_1^{\dagger}A_1-A_2A_2^{\dagger}\nonumber\eeqr The
Stokes operators obey similar commutation relations
$[S_i,S_j]=2\epsilon_{ijk}S_k$ ($i=1,2,3$) similar to the atomic spin and
therefore provide a useful and intuitive representation of the quantum state of
the field in our situation. Since we assumed $\langle A_2\rangle=0$, the Stokes vector is
parallel to the atomic spin: $\langle S_z\rangle=\langle
A_1\rangle^2$ and $\langle S_x\rangle=\langle S_y\rangle=0$. Let
us assume that the incident vacuum is squeezed for the amplitude
quadrature $A_p$ and that the squeezing
bandwidth is broad with respect to the cavity bandwidth, so
that its minimal noise spectrum is $\langle (\delta
A_p^{in})^2\rangle=e^{-2r}$. As $\delta S_x=-\langle A_1\rangle
\delta A_p$, the field is also said to be $S_x$-polarization squeezed.

It is easy to see
that the first terms in the r.h.s of (\ref{jxeit}-\ref{jyeit}) derive from
an effective Hamiltonian
\beqr H_{E}=-\hbar\frac{
2g^2}{\gamma(1+2C)\sqrt{T}}\left[J_xS_y^{in}-J_yS_x^{in}\right]\lb{HE}\eeqr
The Langevin forces in (\ref{jxeit}-\ref{jyeit}) being white noises, their contribution to the
atomic noise is the same for any component in the ($x,y$)-plane.
By looking at (\ref{jxeit}-\ref{jyeit}), one can see that,
for a $S_x$-squeezed incident field, the least noisy spin component
will be the $x$-component. Its normalized variance is
\beqr\lb{var} \Delta J_{min}^2&=&\frac{1}{|\langle J_z\rangle|/2}
\left(\frac{1}{2\pi}\int d\omega
\langle \delta J^2_x(\omega)\rangle\right)\\
\nonumber&=&\frac{2C}{1+2C}\frac{\Gamma_E}{(1+2C)\tilde{\gamma}_0}e^{-2r}
+\frac{\Gamma_E}{(1+2C)^2\tilde{\gamma}_0}+
\frac{\gamma_0}{\tilde{\gamma}_0}\eeqr
We used the fact that $\langle
f_x(\omega)f_x(\omega')\rangle=2\pi\delta(\omega+\omega')\;N\gamma_0/2$
and $\langle
F_y(\omega)F_y(\omega')\rangle=2\pi\delta(\omega+\omega')\;N\gamma/2$. The three terms in
(\ref{var}) can be understood as the coupling with the incident field ($\propto e^{-2r}$), the noise contribution
of the optical dipole ($\propto \Gamma_E$) and the noise due to the loss of coherence in the ground state
($\propto \gamma_0$), respectively.
We characterize the transfer efficiency as the ratio of the atomic squeezing created in the
ground state to the
incident field squeezing
\beqr \eta &\equiv &\frac{1-\Delta
J_{min}^2}{1-e^{-2r}},\nonumber\eeqr
perfect transfer corresponding to $\eta=1$. In an ideal EIT configuration and in the lower frequency
approximation, this parameter thus takes the form
\beqr
\eta_E=\frac{2C}{1+2C}\frac{\Gamma_E/(1+2C)}{\gamma_0+\Gamma_E/(1+2C)}\lb{etaeit}\eeqr
The transfer is almost perfect - $\eta_E\sim 1$ - for a good
cooperative behavior ($C\gg 1$) and when the effective EIT pumping
is much larger than the loss rate in the ground state
[$\Gamma_E/(1+2C)\gg\gamma_0$]. Note that, for a closed system
($\gamma_0=0$), the efficiency takes the extremely simple form
\beqr \eta_{max}=\frac{2C}{1+2C}\sim
\frac{\text{Coupling}}{\text{Coupling+Atomic Noise}}\nonumber\eeqr
which emphasizes the central role played by the cooperativity to
quantify the atom/field interaction in cavity. The noise degrading
the transfer [$\propto 1/(1+2C)$] can thus be made very small with respect to the coupling
[$\propto 2C/(1+2C)$] by increasing the
cooperativity, i.e. for large atomic samples ($C\propto N$).
In a cavity configuration, the cooperativity easily reaches 100-1000,
ensuring in principle a perfect transfer.

\subsection{Analogy with the Raman configuration}\lb{ramaneit}

In a previous work \ct{dantan1}, we studied squeezing transfer in
a $\Lambda$ system in the case where the fields are strongly
detuned with respect to the atomic resonance
($\Delta_{1,2}\gg\gamma$). In such a configuration the three-level
system can be reduced to an effective two-level system for the ground state. We
denote the Raman optical pumping rate by
$\Gamma_R=\gamma\Omega^2/\Delta^2$. When the effective two-photon detuning
$\tilde{\delta}=\delta+\Omega^2/\Delta$, as well as the effective cavity
detuning $\tilde{\Delta}_c=\Delta_c-g^2N/\Delta\tau$ are cancelled,
the equations for the $x,y$-spin components read
\beqr
(\tilde{\gamma}_0-i\omega)\djx&=&+\frac{g^2N}{\Delta\sqrt{T}}\dsy+\tilde{f}_x\lb{jxraman}\\
(\tilde{\gamma}_0-i\omega)\djy&=&-\frac{g^2N}{\Delta\sqrt{T}}\dsx+\tilde{f}_y\lb{jyraman}\eeqr
with $\tilde{\gamma}_0=\gamma_0+(1+2C)\Gamma_R$ and
\beqr\tilde{f}_{x}=
f_{x}-\frac{\Omega}{\Delta}F_{y},\hspace{0.3cm}\tilde{f}_{y}=
f_{y}+\frac{\Omega}{\Delta}F_{x}\lb{framan}\eeqr
These equations were derived from the effective equations given in
\ct{dantan1} by eliminating the intracavity field and introducing
the incident Stokes vector as in the previous Section. As in EIT,
one can deduce an effective Raman Hamiltonian
\beqr H_R=\hbar\frac{
2g^2}{\Delta\sqrt{T}}\left[J_xS_x^{in}+J_yS_y^{in}\right]\lb{HR}\eeqr
Assuming again a $S_x$-squeezed incident field, the minimal
variance is now that of the $y$-component, and one gets the
following efficiency
\beqr\eta_R=\frac{2C}{1+2C}\frac{(1+2C)\Gamma_R}{\gamma_0+(1+2C)\Gamma_R}\lb{etaraman}\eeqr
The similarity between the EIT and Raman configuration appears
clearly by comparing
(\ref{jxeit}-\ref{jyeit}-\ref{feit}-\ref{etaeit}) to
(\ref{jxraman}-\ref{jyraman}-\ref{framan}-\ref{etaraman}). The
equations are formally identical by making the substitution
\beqr (1+2C)\gamma\longleftrightarrow\Delta\lb{substitution}\eeqr
The important result is that the transfer efficiency takes the
same form in both the on-resonant and strongly off-resonant
situations

\beqr \eta&=&\frac{2C}{1+2C}\frac{\Gamma}{\gamma_0+\Gamma}\lb{etageneral}\\
&&\text{with}\hspace{0.3cm}\Gamma=\frac{\Gamma_E}{1+2C}\hspace{0.3cm}\text{or}\hspace{0.3cm}(1+2C)\Gamma_R\nonumber\eeqr

The effective pumping rate, $\Gamma=\Gamma_E/(1+2C)$ or $(1+2C)\Gamma_R$, is obtained in each case
by making the substitution (\ref{substitution}), and can be made much larger than $\gamma_0$
with an adequate choice of $\Omega$.
Note, however, that the EIT and Raman Hamiltonian are identical to
a spin rotation by $\pi/2$ in the $(x,y)$-plane. We retrieve a well-known "$\pi/2$" phase-shift phenomenon when
going from "on-resonance" to "off-resonance".

\subsection{Transfer for an arbitrary detuning}\lb{Delta}

\begin{figure}
  \centering
  \includegraphics[width=8cm]{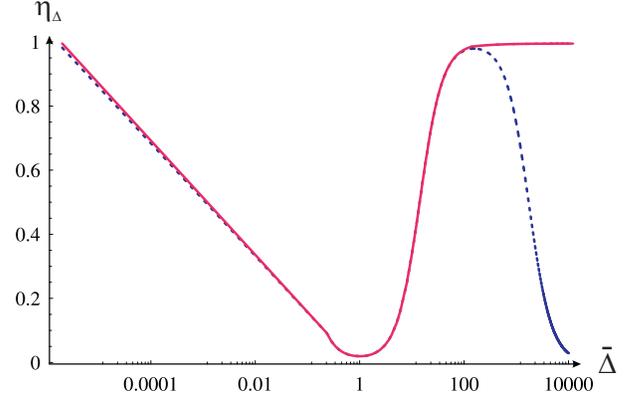}
  \caption{Transfer efficiency versus $\bar{\Delta}$ for
  $\gamma_0=0$ [plain] and $\gamma_0=\gamma/1000$ [dash] ($C=100$,
  $\gamma_E=15$).}\label{vardelta}
\end{figure}

The predictions given by the low frequency approximation in both
the EIT and Raman configurations could lead one to expect squeezing
transfer for any value of the one-photon detuning $\Delta$,
provided one maintains the optimal transfer conditions
$\tilde{\Delta}_c=\tilde{\delta}=0$. Moreover, given the $\pi/2$
rotation of the squeezed spin component when going over from
on-resonance to off-resonance, one expects the squeezed component
to continuously rotate from $0$ to $\pi/2$ when the detuning is
increased.\\
Using (\ref{dpr}-\ref{da}-\ref{generalLF}) one finds the optimal
transfer conditions to be
\beqr
\tilde{\Delta}_c&=&\Delta_c-2C\kappa\frac{\gamma\Delta}{\gamma^2+\Delta}=0\lb{deltac}\\
\tilde{\delta}&=&\delta+
\Gamma_E\frac{\gamma\Delta^3+(1-2C)\gamma^3\Delta}{(\gamma^2+\Delta^2)[(1+2C)\gamma^2+\Delta^2]}=0\lb{deltabar}\eeqr
Eq. (\ref{generalLF}) then leads to the general equation for the spin
component
$J_{\theta}$ with
angle $\theta$ in the $(x,y)$ plane
\beqr (\tilde{\gamma}_0-i\omega)\delta J_{\theta}&=&\alpha \delta
A^{in}(\theta-\phi)\\
\nonumber
&&+\beta[e^{-i(\theta-\phi')}F_2+e^{i(\theta-\phi')}F_2^{\dagger}]/2\\
\nonumber
&&+[e^{-i\theta}f_r+e^{i\theta}f_r^{\dagger}]/2\eeqr
with $\alpha$, $\beta$, $\tilde{\gamma}_0$, $\phi$ and $\phi'$ functions depending on $\Delta$.
Starting again with a $S_x$-squeezed field, the squeezed spin component
will be $J_{\phi}$. After straightforward
calculations the optimal efficiency for a given $\Delta$ is
\beqr \eta_{\Delta}&=&\frac{2C\gamma_E(1+\bar{\Delta}^2)^2}
{(1+2C+\bar{\Delta}^2)}\times\\
\nonumber&&\frac{1}{\sigma(1+\bar{\Delta}^2)(1+2C+\bar{\Delta}^2)+\gamma_E(1+(1+2C)\bar{\Delta}^2)}\eeqr
with $\sigma=\gamma_0/\gamma$, $\gamma_E=\Gamma_E/\gamma$ and $\bar{\Delta}=\Delta/\gamma$.
This efficiency is plotted in Fig. \ref{vardelta} for the two
cases considered previously: $\gamma_0=0$ and $\gamma_0\simeq 0$.
In the first case the efficiency is optimal in EIT
($\bar{\Delta}=0$, $\eta_{\Delta}=\eta_{max}$), decreases to a minimum for $|\bar{\Delta}|=1$
($\eta\sim 2/C\ll 1$) and increases again back to its maximal
value $\eta_{max}$ when $\bar{\Delta}\gg 1$. The squeezed component
angle can be shown to be $\theta_{sq}=\text{Arctan}\;\bar{\Delta}$, which varies as
expected by $\pi/2$ when $\bar{\Delta}$ goes from 0 to infinity. One retrieves that the
transfer is optimal either in an EIT or a Raman configuration. However, the
transfer is really degraded in the intermediate
regime $\bar{\Delta}\sim 1$.\\
If one takes into account losses in the ground state ($\gamma_0\neq 0$), the efficiency
now reaches a maximum for $\bar{\Delta}\gg 1$
\beqr \eta_{\Delta}&\simeq& \eta_{max}\left(1-2\sqrt{\frac{\gamma_0}{\Gamma_E}}\right)
\hspace{0.5cm}(C\gg 1,\;\gamma_0\ll \Gamma_E)\nonumber\\
\nonumber\text{for}&&\bar{\Delta}\simeq\sqrt{2C}\left(\frac{\Gamma_E}{\gamma_0}\right)^{1/4}\eeqr
before decreasing when the coupling in $(1+2C)\Gamma_R$ becomes too small as $\bar{\Delta}$ is increased ($\Omega$ being fixed)
to compensate for the noise associated to the loss of coherence
$\gamma_0$ [see Eq. (\ref{etaraman})]. These effects stress the fragility of
the squeezing transfer with respect to dissipation and explain
why dissipation-less situations like EIT or Raman are favorable.

\section{Full three-level calculation}\lb{full}

From the Heisenberg-Langevin equations given at the beginning we
calculated without approximation the spin covariance matrix and now
compare it with the analytical model used in the previous Sections.

\subsection{Exact calculation in EIT}\lb{exact}

\begin{figure}
  \centering
  \includegraphics[width=9cm]{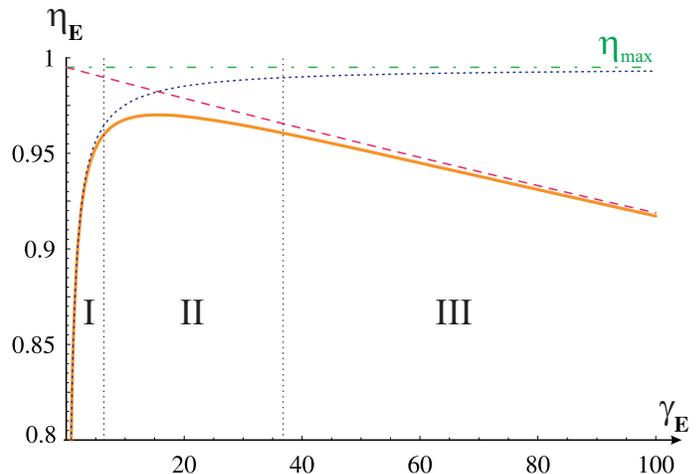}
  \caption{EIT Transfer efficiency versus EIT pumping rate: analytical (\ref{etaeitexact}) [plain],
  low frequency approximation (\ref{etaeit}) [dots] and
  lossless system (\ref{etaezero}) [dashed]. The three regimes are I ($\Gamma\ll\gamma_0$),
  II ($\gamma_0\ll\Gamma\ll\gamma,\kappa$) and III ($\Gamma\gtrsim\gamma,\kappa$).
  Parameters: $C=100$, $\sigma=1/1000$, $\rho=1/2$. The optimal pumping rate is then $\gamma_E^*\simeq 15$.}\label{etatotal}
\end{figure}

In the previous sections we neglected the frequencies larger than
the atomic fluctuations evolution constant $\tilde{\gamma}_0$,
assuming that $\kappa,\gamma\gg\tilde{\gamma}_0$. We therefore
neglected high atom-field coupling frequencies due to the cavity.
However, the analytical calculation of the minimal spin variance
in EIT is possible using the Fourier transforms of
(\ref{dpr}-\ref{dpd}-\ref{da}). In EIT
($\Delta=\tilde{\Delta}_c=\tilde{\delta}=0$), the resulting
equation for the $x$-component reads
\beqr\nonumber
\left[\gamma_0-i\omega+\frac{\Omega^2(\kappa-i\omega)}{D(\omega)}\right]\djx=&\\
\nonumber
+\frac{g^2N}{D(\omega)}\sqrt{\frac{2\kappa}{\tau}}\dsx+f_x-&\frac{\Omega(\kappa-i\omega)}{D(\omega)}F_y\\
\text{with}\hspace{0.3cm}
D(\omega)=(\kappa-i\omega)(\gamma-i\omega)+\frac{g^2N}{\tau}&\nonumber
\eeqr
If the incident field is
$S_x$-squeezed, we know the $x$-component will be squeezed.
However, a well-know coupling frequency
($\omega_c\simeq \sqrt{2C/\rho}\gamma$) appears at high frequency \ct{vernac1},
resulting in an increase of atomic noise, and, consequently, in a
degradation of the atomic squeezing. After integration, the exact efficiency is
\beqr
\eta_E=&&\frac{2C\gamma_E}{(1+2C)\sigma+\gamma_E}\lb{etaeitexact}\\
\nonumber
&\times&\frac{1+\rho+\sigma\rho}{2C(1+\rho)+(1+\sigma)(1+\rho+\sigma\rho+\sigma\rho^2+\gamma_E\rho^2)}\eeqr
with $\gamma_E=\Gamma_E/\gamma$, $\rho=\gamma/\kappa$ and
$\sigma=\gamma_0/\gamma$. Three regimes can be distinguished: for
very small values of the effective pumping
$\Gamma=\Gamma_E/(1+2C)$ compared to the loss rate in the
ground state $\gamma_0$, one retrieves the low frequency
(\ref{etaeit}) result as can be seen from Fig. \ref{etatotal}: the
efficiency is bad as long as the loss of coherence in the
ground state $\gamma_0$ is not overcome by the pumping. In an
intermediate regime $\gamma_0\ll\Gamma\ll\gamma,\kappa$ the
efficiency reaches its maximum. The optimal pumping rate $\gamma_E^*$ can be shown to
be proportional to $\sqrt{\sigma}$
\beqr \nonumber\frac{\gamma_E^*}{1+2C}\simeq
\frac{\sqrt{1+\rho}}{\rho}\sqrt{\sigma}\hspace{0.5cm}(C\gg 1,
\sigma\ll 1)\eeqr
in good agreement with the results shown in Fig. \ref{etatotal}.
For values of $\Gamma$ comparable to $\gamma$, $\kappa$, the
efficiency is no longer well reproduced by the low frequency
approximation, since the adiabatical eliminations are no longer
valid. In this regime, the efficiency asymptotically reaches that
of a closed system ($\gamma_0=0$), for which (\ref{etaeitexact})
reduces to a monotonously decreasing function of $\gamma_E$
\beqr
\eta_E^0=\frac{2C}{1+2C+\gamma_E\frac{\rho^2}{1+\rho}}\lb{etaezero}\eeqr
The optimal transfer is naturally obtained by making a compromise between the
coupling and the atomic noise, and occurs in the intermediate regime II between
regime I, for which the coupling is small and the atomic noise due to ground state coherence losses dominates, and
regime III, in which the coupling is large, but the atomic noise due to spontaneous emission is more important.

\subsection{Robustness with respect to two-photon detuning}

\begin{figure}[h]
  \centering
  \includegraphics[width=8cm]{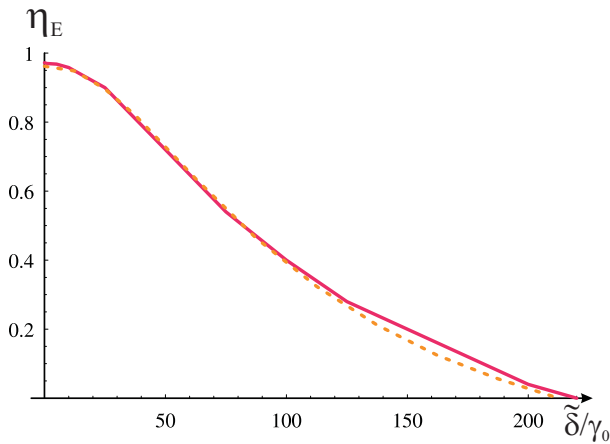}
  \caption{Transfer efficiency versus $\tilde{\delta}/\gamma_0$ in EIT [plain] and Raman [dash] schemes. Parameters:
  $C=100$, $\sigma=1/1000$, $e^{-2r}=0.5$. With these values,
  $\Gamma_E$ and $\Gamma_R$ are chosen so that, in both cases, $\tilde{\gamma}_0\simeq75\gamma_0$.}\label{vardeltatilde}
\end{figure}

In a $\Lambda$ scheme, the coherence created between the
ground state sublevels strongly depends on the two-photon resonance, the
width of which is given by the effective atomic decay constant
$\tilde{\gamma}_0$. In Fig. \ref{vardeltatilde} we plot the
transfer efficiency for the least noisy spin component as a
function of the two-photon detuning for a zero-cavity detuning, that is, when
(\ref{deltac}) is fulfilled, but not (\ref{deltabar}).
In addition to rotating the maximally squeezed component in the
$(x,y)$-plane, the spin squeezing is clearly destroyed as soon as
$\tilde{\delta}\sim\tilde{\gamma}_0$. We would like to emphasize that both
EIT and Raman configurations are equally sensitive to this
two-photon resonance condition. This similarity adds to the
resemblance already stressed in Sec. \ref{ramaneit}.

\subsection{Transformation to the "$\langle A_2\rangle=0$" basis}

In this Section we show that any incident field state can actually
be transferred to the atoms in EIT. To simplify the discussion let
us assume again that the modes $A_{1,2}$ interacting with the
transitions of the $\Lambda$ system are orthogonally polarized
modes. Because of the similarities existing between the Stokes
vector and the atomic spin, the results obtained in the special
case $\langle A_2\rangle=0$ and $\langle J_z\rangle=N/2$
considered previously can be applied to any polarization state of
the incident field. The Hamiltonian for a $\Lambda$ system in EIT
reads
\beqr \nonumber H=\hbar
[g_1A_1^{\dagger}P_1+g_2A_2^{\dagger}P_2+\text{h.c.}]\eeqr
If both $\langle A_1\rangle$ and $\langle A_2\rangle$ are
non-zero, one can always turn to a basis $(A_1',A_2')$ for which
$\langle A_2'\rangle=0$ via a rotation $\textrm{R}$ in the Poincar\'{e} sphere.
The Hamiltonian is invariant under the same rotation performed on
the atomic spin, the atoms are pumped into the dark state $|D\rangle=\textrm{R}|2\rangle$
and the $A'_2$ field state will thus imprint on the
atomic spin. Let us assume, for instance, that $A_1$ and $A_2$ have
minimal noises $e^{-2r_{i}}$ for the same quadratures and
take $\Omega_i=g_i\langle A_i\rangle$ ($i=1,2)$ as real numbers. The dark state is then
\beqr \nonumber |D\rangle=\frac{-\Omega_2|1\rangle+\Omega_1|2\rangle}{\sqrt{\Omega_1^2+\Omega_2^2}}\eeqr
The minimal atomic variance is then a weight of the $A_{1,2}$ modes
squeezings
\beqr \nonumber\Delta J^2_{min}\simeq\frac{\Omega_2^2
e^{-2r_1}+\Omega_1^2 e^{-2r_2}}{\Omega_1^2+\Omega_2^2}\eeqr
One finds naturally that one cannot transfer more than the
squeezing of one mode.

\section{Reading scheme and quantum memory}\lb{lecture}

We have shown how the quantum state of the incident field could be
transferred to the atomic spin in the ground state. Note that the
lifetime associated to the ground state is quite long for cold
atoms, and therefore the quantum information can be stored for a
long time (several ms).  Let us start with our spin squeezed atomic
ensemble and switch off the fields at time $t=0$.
The spin squeezing then decreases on a time scale given by the ground state
lifetime $\gamma_0^{-1}$. After a variable delay, we rapidly switch on again the pump field,
field $A_2^{in}$ being in a coherent vacuum state, and we
look at the fluctuations of the field exiting the cavity $A^{out}_2=\sqrt{T}A_2-A_2^{in}$.
Let us assume an EIT configuration for simplicity and
start with a $J_x$-squeezed atomic spin; one expects its fluctuations to
imprint on the $S_x$ component of the outgoing field [see (\ref{HE})].

\subsection{Standard homodyne detection}

We assume a standard homodyne detection scheme with a constant local oscillator and calculate the noise power of $A^{out}_q$
measured by a spectrum analyzer integrating
during a time $T_0$ over a frequency bandwidth $\Delta\omega$ centered around zero-frequency
\beqr P(t)=\int_{-\frac{\Delta\omega}{2}}^{\frac{\Delta\omega}{2}}\frac{d\omega}{T_0}\int_{t}^{t+T_0}d\tau \int_{t}^{t+T_0}d\tau'
e^{-i\omega(\tau-\tau')}\mathcal{C}(\tau,\tau')\nonumber\eeqr
where $\mathcal{C}(\tau,\tau')=\langle \delta A_p^{out}(\tau)\delta A_p^{out}(\tau')\rangle$ is the correlation function of $A_p^{out}$.
Note that $T_0$ and $\Delta\omega$ must satisfy $T_0\Delta\omega\geq 2\pi$.
In the low frequency approximation and in the "good" regime for transfer ($\gamma_0\ll\Gamma_E/(1+2C)\ll \gamma,\kappa$; see Sec. \ref{exact}),
the correlation function of $A_p^{out}$ may be calculated via
the Laplace-transforms of Eqs. (\ref{dpr}-\ref{dpd}-\ref{da})
\beqr \mathcal{C}(\tau,\tau')=\delta(\tau-\tau')-\frac{4C\Gamma_E}{(1+2C)^2}R_{at}e^{-\tilde{\gamma}_0(\tau+\tau')}\label{correlation}\eeqr
where $R_{at}=1-\Delta J^2_{min}$ represents the atomic squeezing when the pump field is switched on again. After some algebra, one gets
\beqr \frac{1}{\Delta\omega}P(t)=1-S(a,b)R_{at} e^{-2\tilde{\gamma}_0t}\nonumber\eeqr
where $S$
is an integral depending on two dimensionless parameters: $a=T_0\tilde{\gamma}_0$ and $b=\Delta\omega/\tilde{\gamma}_0$,
which, for large values of $C$, is equal to
\beqr S(a,b)=2\int_{-b/2}^{b/2}
\frac{d\bar{\omega}}{ab}\frac{1+e^{-2a}-2e^{-a}\cos(\bar{\omega}a)}{1+\bar{\omega}^2}\nonumber\eeqr
with $\bar{\omega}=\omega/\tilde{\gamma}_0$. This integral, which can be understood as the signal-to-noise ratio of the readout process,
is also the ratio of the measured field squeezing $R_{out}=1-P(0)/\Delta\omega$ to the initial atomic squeezing $R_{at}$.
Squeezing may thus be transferred back from the atoms to the field. This squeezing decreases back to a coherent vacuum state on a time scale
$\tilde{\gamma}_0^{-1}$ given by the atoms.
$S(a,b)$ is optimal when the spectrum analyzer is
Fourier-limited: $b=2\pi/a$, and when the time measurement is of the order of the inverse of the atomic spectrum width: $a\simeq 1.3$.
Under these conditions, the integral is about 0.64, and about two-third of the atomic squeezing is transferred
to the field exiting the cavity: $R_{out}\simeq 0.64 R_{at}$.

\subsection{Temporal matching}

This imperfect readout comes from the fact that the local oscillator detecting the
fluctuations of vacuum mode exiting the cavity is not perfectly matched with the atomic squeezing spectrum \ct{molmer}.
It is possible to reach a
perfect readout by choosing the right temporal profile for the local oscillator: $E_{LO}(\tau)=e^{-\zeta\tilde{\gamma}_0\tau}$,
with $\zeta$ a dimensionless adjustable parameter.
The spectrum analyzer now measures \beqr P(t)=\int_{-\frac{\Delta\omega}{2}}^{\frac{\Delta\omega}{2}}\frac{d\omega}{T_0}&&
\int_{t}^{t+T_0}d\tau \int_{t}^{t+T_0}d\tau'
e^{-i\omega(\tau-\tau')}\nonumber\\
&&\times E_{LO}(\tau)E_{LO}(\tau')\mathcal{C}(\tau,\tau')\nonumber\eeqr
Using the correlation function (\ref{correlation}), one gets
\beqr \nonumber\frac{1}{\Delta\omega}P(t)&=&N(a,\zeta)-S(a,b,\zeta)e^{-2\tilde{\gamma}_0t}R_{at}\\
\nonumber\textrm{with}&&N(a,\zeta)=\frac{1-e^{-2\zeta a}}{2\zeta a}\\
\nonumber S(a,b,\zeta)&=&\frac{2}{ab}\int_{-b/2}^{b/2}d\bar{\omega}\frac{1+e^{-2a(1+\zeta)}-2e^{-a(1+\zeta)}\cos(a\bar{\omega})}{(1+\zeta)^2+\bar{\omega}^2}\eeqr
$N(a,\zeta)$ represents the noise level in the absence of atomic squeezing and $S(a,b,\zeta)$ the amplitude of the atomic squeezing transferred to
the field. The field squeezing can be expressed as \beqr \nonumber R_{out}(t)\equiv 1-\frac{P(t)}{\Delta\omega N(a,\zeta)}=
\frac{S(a,b,\zeta)}{N(a,\zeta)}e^{-2\tilde{\gamma}_0t}R_{at}\eeqr
and, for short times, the readout efficiency is $\mu\equiv R_{out}(0)/R_{at}=S(a,b,\zeta)/
N(a,\zeta)$. This efficiency is optimized when the spectrum analyzer is Fourier limited and when the integration time is larger than the inverse of the
atomic spectrum width: $b=2\pi/a$ and $a\gg 1$. In this case, one has $N(a,\zeta)\sim 1/(2a\zeta)$ and $S(a,b,\zeta)\sim 2/a(1+\zeta)^2$,
so that the efficiency reads \beqr \mu\sim\frac{4\zeta}{(1+\zeta)^2}\nonumber\eeqr It is maximal and equal to 1 when $\zeta=1$, i.e. when the temporal
profile of the local oscillator perfectly matches the atomic noise spectrum. It is thus possible to fully retrieve the atomic squeezing stored into
the atoms in an EIT configuration. Note that the same results can be obtained
in a Raman configuration, in the regime $\gamma_0\ll(1+2C)\Gamma_R\ll\gamma,\kappa$. In both schemes,
to ensure that the retrieved squeezing indeed originates from the atoms, one may vary the delay between the switching on and off, and check the
exponential decay of the squeezing with $\gamma_0$.

\section{Conclusion}

To conclude, we have shown that a quasi-ideal squeezing transfer should
be possible between a broadband squeezed vacuum and the ground state spin of $\Lambda$-type atoms.
The cavity interaction allows for good transfer
Our results for a cavity configuration are consistent with those obtained in single pass schemes with
thick atomic ensembles \ct{lukin,polzik,kuzmich,kozhekin,molmer},
although efforts are still being conducted to develop a full free-space quantum treatment \ct{duan2}.
The most favorable schemes are those minimizing dissipation, such as EIT or Raman \ct{dantan1},
for the
fluctuations of the intracavity field imprint on the atomic spin, thus mapping the incident
field state onto the atoms. The relevant physical parameter for the transfer efficiency is the cooperativity,
which quantifies the collective spin-field interaction and which can be large in a cavity scheme.
The mapping efficiency was evaluated taking into account possible
losses in the ground state. Its robustness with respect to a detuning from the two-photon resonance
is shown to be the same in EIT and in the Raman scheme.
We also generalized the EIT results to the case in which both fields have non-zero intensity. The atomic squeezing is
in this case a combination of the incident field squeezings. This is related to the fact that, in EIT, the atoms are
pumped into a dark-state and the atomic medium is then transparent for a certain combination of the fields.
Such a dark-state pumping was exploited for double-$\Lambda$ atoms in \ct{dantan2} to generate "self spin-squeezing"
using only coherent fields.

Last, we propose a simple reading scheme for the atomic state,
allowing a quantum memory-type operation. When the pump field is switched back on, the outgoing vacuum is squeezed by the atoms,
and the atomic squeezing can be fully transferred back by
temporally matching the local oscillator used to detect the outgoing vacuum fluctuations with the atomic spectrum.
To our knowledge, it is the first instance in which the conservation of quantum variables is predicted in an EIT scheme
within a full quantum model.


\bigskip

\begin{thebibliography}{99}

\bibitem{divincenzo} D.P. Di Vincenzo, Fortschr. Phys.
\textbf{48}, 771 (2000).

\bibitem{lukin} M.D. Lukin, S.F. Yelin, and M. Fleischhauer, Phys.
Rev. Lett. \textbf{84}, 4232 (2000); M.D. Lukin and M.
Fleischhauer, Phys. Rev. Lett. \textbf{84}, 5094 (2000); M.
Fleischhauer and M.D. Lukin, Phys. Rev. A \textbf{65}, 022314
(2002). For a review, see M.D. Lukin, Rev. Mod. Phys. \textbf{75},
457 (2003).

\bibitem {polzik}J. Hald, J.L. S\o rensen, C. Schori, and
E.S. Polzik, Phys. Rev. Lett. \textbf{83}, 1319 (1999); B.
Julsgaard, A. Kozhekin, and E.S. Polzik, Nature \textbf{413}, 400
(2001).

\bibitem{duan} L.M. Duan, J.I. Cirac, P. Zoller, and E.S. Polzik, Phys. Rev.
Lett. \textbf{85}, 5643 (2000);
L.M. Duan, M.D. Lukin, J.I. Cirac, and P. Zoller,
Nature \textbf{414}, 413 (2001).

\bibitem {hau}S.E. Harris and L.V. Hau, Phys. Rev. Lett. \textbf{82}, 4611
(1999); C. Liu, Z. Dutton, C.H. Behroozi, and L.V. Hau, Nature
\textbf{409}, 490 (2001).

\bibitem {phillips} D.F. Phillips, A. Fleischhauer, A. Mair, and R.L.
Walsworth, and M.D. Lukin, Phys. Rev. Lett. \textbf{86}, 783
(2001).

\bibitem{harris} S.E. Harris, Phys. Today \textbf{50}, 36 (1997).

\bibitem{polzik2} C. Schori, B. Julsgaard, J.L S\o rensen, and
E.S. Polzik, Phys. Rev. Lett. \textbf{89}, 057903 (2002).

\bibitem {wineland} D.J. Wineland, J.J. Bollinger, W.M. Itano, and D.J. Heinzen,
Phys. Rev. A \textbf{50}, 67 (1994).

\bibitem{polzik3} A. Kuzmich, K. M\o lmer, and E.S. Polzik, Phys. Rev. Lett.
\textbf{79}, 4782 (1997).

\bibitem{kuzmich} A. Kuzmich, L. Mandel, and N.P. Bigelow, Phys. Rev. Lett. \textbf{85}, 1594 (2000).

\bibitem {vernac1}L. Vernac, M. Pinard, and E. Giacobino, Phys. Rev. A
\textbf{62}, 063812 (2000); L. Vernac, M. Pinard, and E.
Giacobino, Eur. Phys. J. D \textbf{17},125 (2001).

\bibitem{kozhekin} A.E. Kozhekin, K. M\o lmer, and E.S. Polzik,
Phys. Rev. A \textbf{62}, 033809 (2000).

\bibitem{molmer} U.V. Poulsen and K. M\o lmer, Phys. Rev. Lett.
\textbf{87}, 123601 (2001).

\bibitem {dantan1} A. Dantan, M. Pinard, V. Josse, N. Nayak, and P.R. Berman,
Phys. Rev. A \textbf{67}, 045801 (2003).

\bibitem{ueda} M. Kitawaga and M. Ueda, Phys. Rev. A \textbf{47},
5138 (1993).

\bibitem{duan2} L.M. Duan, J.I. Cirac, and P. Zoller, Phys. Rev. A \textbf{66}, 023818 (2002).

\bibitem{dantan2} A. Dantan, M. Pinard, and P.R. Berman, Eur.
Phys. J. D \textbf{27}, 193 (2003).

\end{thebibliography}
\end{document}